\documentclass[preprint]{ptephy_v1}

\pdfoutput=1
\usepackage{graphicx}

\usepackage{lineno}

\newcommand{\klpippim}{K_L \to \pi^+ \pi^-}
\newcommand{\klpiopio}{K_L \to \pi^0 \pi^0}
\newcommand{\kspippim}{K_S \to \pi^+ \pi^-}
\newcommand{\kspiopio}{K_S \to \pi^0 \pi^0}
\newcommand{\pippim}{\pi^+\pi^-}
\newcommand{\piopio}{\pi^0\pi^0}
\newcommand{\ko}{K^0}
\newcommand{\kobar}{\overline{K}^0}
\newcommand{\keven}{K_\mathrm{even}}
\newcommand{\kodd}{K_\mathrm{odd}}
\newcommand{\reepoe}{Re(\epsilon'/\epsilon)}

\newcommand{\NKLDecay}{N_{K_L\ \mathrm{decay}}}

\newcommand{\Nobs}[1]{N_{#1}}
\newcommand{\Acc}[1]{A_{#1}}

\newcommand{\kppipnn}{K^+ \to \pi^+ \nu \overline{\nu}}
\newcommand{\klpionn}{K_L \to \pi^0 \nu \overline{\nu}}
\newcommand{\kpinn}{K \to \pi\nu\overline{\nu}}
\newcommand{\pionn}{\pi^0 \nu \overline{\nu}}
\newcommand{\kppippio}{K^+ \to \pi^+ \pi^0}
\newcommand{\kmutwo}{K^+ \to \mu^+ \nu}

\newcommand{\bra}[1]{\langle #1|}
\newcommand{\ket}[1]{|#1\rangle}
\newcommand{\kokobarvec}{
	\begin{pmatrix}
		\ko(t)\\
		\kobar(t)
	\end{pmatrix}
}

\newcommand{\cp}{$CP$}
\newcommand{\sm}{Standard Model}
\newcommand{\superweakModel}{the Superweak model}

\begin{document}

\title{Kaon Experiments}



\author{Taku Yamanaka}
\affil{Osaka University\footnote{Current affiliation: KEK}\email{taku@post.kek.jp}}

%
%


\begin{abstract}%
After \cp\ violation was discovered in the $\klpippim$ decay, 
many theories were proposed to explain it, 
and the Kobayashi-Maskawa model and \superweakModel\ 
lasted for many years as strong candidates.
High-precision kaon experiments with 
many new techniques and improvements  
rejected \superweakModel\ and supported the Kobayashi-Maskawa model in 1990's.
After then, rare kaon decay experiments are studying 
$\kpinn$ decays to search for \cp\ violation caused by
new physics beyond the \sm.
Various techniques have been developed to increase the sensitivity and to suppress
backgrounds.
\end{abstract}

\subjectindex{C02, C03, C30}

\maketitle

\section{Introduction}
After the \cp\ violation was discovered in the $\klpippim$ decay in 1964\cite{cronin},
the $\klpiopio$ decay was also observed\cite{banner1968},
and
many theories were proposed to explain the phenomena.
Among them, the Kobayashi-Maskawa model with three generations of quarks\cite{km} and \superweakModel\cite{sw}
remained as strong candidates, but
in 1970's, it seemed almost impossible to do an experiment 
to test which is correct.
However, now the Kobayashi-Maskawa model has been established and it is a part of 
the \sm.
What made this possible?

In this article, we will review how kaon experiments have contributed 
to establish the Kobayashi-Maskawa model, and how kaon experiments are
trying to search for \cp\ violation caused by physics beyond the \sm.

\section{Where did the \cp\ violation come from?}
\subsection{What was known back in 1973?}
By 1973, when the paper by Kobayashi and Maskawa was published,
a charge asymmetry in \(K_L \to \pi^\pm e^\mp \nu\) had 
also been observed\cite{ke3delta},
proving that in $K_L$,
the amplitude of $\ko$ is slightly larger than the amplitude of $\kobar$.%
\footnote{The initial state can be tagged by the lepton charge, 
as \(\ko \to \pi^-e^+\nu\) and \(\kobar \to \pi^+e^-\overline{\nu}\).}
The asymmetry also shows that $K_L$ is not a pure \cp-odd state,
\(\ket{\kodd} = (\ket{\ko} - \ket{\kobar})/\sqrt{2}\), 
but it also has a small admixture ($\epsilon$) of \cp-even state,
\(\ket{\keven} = (\ket{\ko} + \ket{\kobar})/\sqrt{2}\), as
\begin{eqnarray}
	\ket{K_L} & \simeq & \frac{1}{\sqrt{2}} 
				\left((1+\epsilon) \ket{\ko} 
					- (1-\epsilon) \ket{\kobar}
				\right)\\
			& = & \ket{\kodd} + \epsilon\ket{\keven} .
\end{eqnarray}
The \cp-violation found in the $K_L \to \pi\pi$ decays was explained 
by the $\keven$ component decaying to the \cp-even $\pi\pi$ state. 
This type of \cp\ violation is called ``indirect \cp\ violation'',
but the origin of $\epsilon$ was unknown.

The fact that $K_L$ stays in this unbalanced equilibrium state means that 
the overall rates of transitions between $\ko$ and $\kobar$ are the same, as
\begin{eqnarray}
	|(1+\epsilon)A(\ko \to \kobar)|^2 & = & |(1-\epsilon)A(\kobar \to \ko)|^2 ,
\end{eqnarray}
where $A(x \to y)$ represents the amplitude for transition $x \to y$.
Because $\epsilon \neq 0$, 
\(|A(\ko \to \kobar)| \neq |A(\kobar \to \ko)|\).
The transition amplitudes are determined by 
the Schr\"odinger equation for the 
two-component wave function of a neutral kaon system,
	\begin{eqnarray}
		i \frac{\partial}{\partial t}\kokobarvec
		& = & H \kokobarvec\\
		& = & \begin{bmatrix}
					M_0 - i\Gamma_0/2		& M_{12} - i\Gamma_{12}/2\\
					M^*_{12} - i\Gamma^*_{12}/2	& M_0 - i\Gamma_0/2
			\end{bmatrix}
			 \kokobarvec ,
%
	\end{eqnarray}
where $K^0(t)$ (\(\overline{K}^0(t)\)) is the amplitude of $\ko$ ($\kobar$) state
at time $t$.
The transition amplitude for $\ko$ returning to the same state, 
\(A(\ko \to \ko)\), is proportional to
\(M_0 - i\Gamma_0/2\).
The $\Gamma_0$ is determined by the sum of amplitudes of 
\(\ko \to f \to \ko\) where $f$ represents the decay final states,
and $M_0$ is determined by the sum of amplitudes of 
\(\ko \to i \to \ko\) where $i$ represents the intermediate virtual states.
The transition amplitude
\(A(\kobar \to \ko)\) is proportional to 
\(M_{12} - i\Gamma_{12}/2\), where
$\Gamma_{12}$ is determined by the sum of amplitudes of 
\(\kobar \to f \to \ko\), and 
$M_{12}$ is determined by the sum of amplitudes of 
\(\kobar \to i \to \ko\),
where $f$ and $i$ are the final and intermediate states, respectively,
which are common to both $\ko$ and $\kobar$.
If, for example, 
$\Gamma_{12}$ is real but $M_{12}$ has an imaginary component\footnote{%
To be more exact, if $\Gamma_{12}$ and $M_{12}$ are not parallel in the complex plane.}, 
then \(|M_{12} - i\Gamma_{12}/2|\) which determines 
\(|A(\kobar \to \ko)|\) would be different from 
 \(|M_{12}^* - i\Gamma_{12}^*/2|\) which determines
\(|A(\ko \to \kobar)|\).
Thus, the indirect \cp\ violation in neutral kaon system can occur 
if there is an imaginary part in $M_{12}$.

\subsection{Source of indirect \cp\ violation}
\begin{figure}[htbp]
	\begin{minipage}{0.45\linewidth}
		\includegraphics[width=\linewidth]{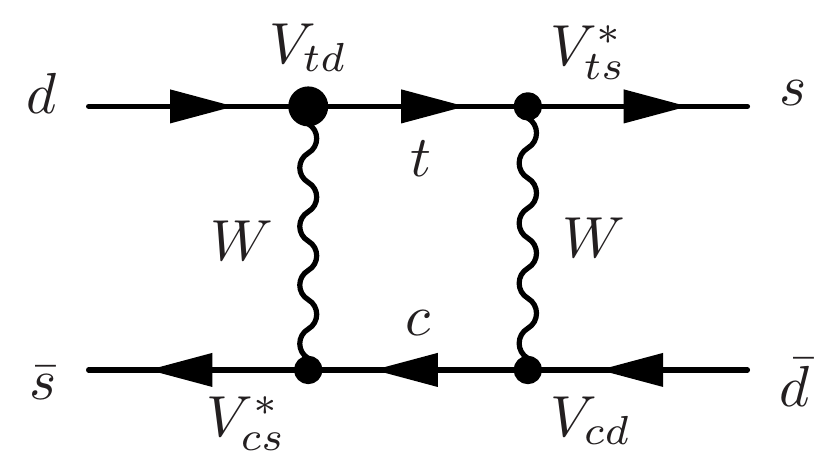}
		\caption{Box diagram for \(\ko \to \kobar\) transition with three generations of quarks which introduces an imaginary part in the amplitude in the Kobayashi-Maskawa model.}
		\label{fig:box_diagram}
	\end{minipage}
	\hspace{0.05\linewidth}
	\begin{minipage}{0.45\linewidth}
            	\includegraphics[width=\linewidth]{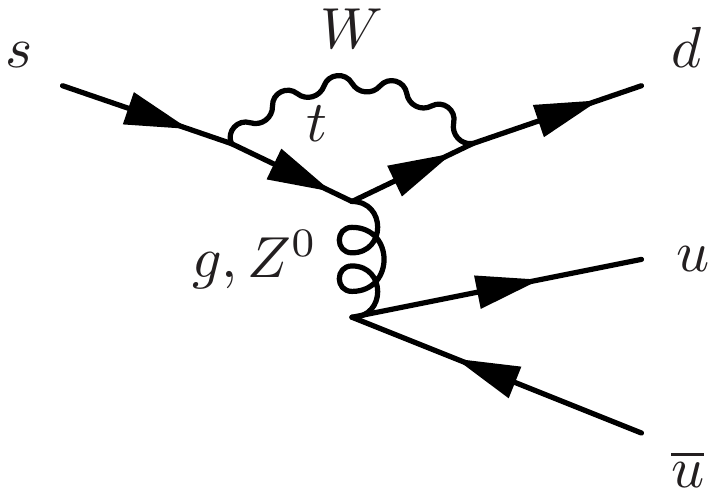}
            	\caption{Penguin diagram for the \(K \to \pi\pi\) decay with 
			three generations of quarks involved.}
            	\label{fig:2pi_penguin}
	\end{minipage}
\end{figure}

What was not known back in 1973 was the source of the imaginary part in
$M_{12}$.
The Superweak model explained that there is a very weak unknown interaction
that changes the strangeness by 2 (\(\kobar \to \ko\)), and that interaction
introduces an imaginary part in $M_{12}$.
Kobayashi and Maskawa explained that 
such an imaginary part
can be naturally introduced by mixings in three generations of quarks.
Figure \ref{fig:box_diagram} shows such an example.
The question then was, which model is correct?
One way to answer it was to check whether the \cp\ is violated in the decay process itself.
The Superweak model cannot violate \cp\ in the decay process 
because it cannot contribute to a \(\Delta S = 1\) process.
However, the Kobayashi-Maskawa model
can naturally introduce an imaginary part in $\Delta S = 1$ decay transition
if three generation of quarks are involved in the decay, such as in a 
penguin diagram shown in Fig.\ref{fig:2pi_penguin}.
Such \cp\ violation in decay processes is called ``direct \cp\ violation'' in
kaon physics.
If one could show that the direct \cp\ violation exists, 
then one could reject \superweakModel\ and support the Kobayashi-Maskwa model.

\section{Is there a direct \cp\ violation?}
One method to test the existence of the direct \cp\ violation was to measure 
the double ratio of branching fractions ($BR$) of four decay modes, 
\begin{eqnarray}
	R & = & \frac{BR(\klpippim)/BR(\kspippim)}
			{BR(\klpiopio)/BR(\kspiopio)}
		\label{eq:R}\\
		& \simeq & 1 + 6 Re(\epsilon'/\epsilon)
\end{eqnarray}
where $\epsilon'$ represents the size of direct \cp\ violation
which is sensitive to the imaginary part of the decay amplitudes of 
\(K^0 \to \pi\pi\).
If $\epsilon'/\epsilon$ is not zero, then it means that there is direct \cp\ violation.%
\footnote{
The basic idea is to use a small difference between
\(\bra{\pippim}H\ket{\kodd}\) and 
\(\bra{\piopio}H\ket{\kodd}\) produced by their isospin dependence.
If there is no direct CP violation, these amplitudes will be zero anyway, and there will be no 
difference between them.
More pedagogical description is available in \cite{bwy}.
}

In early 1970's, existing experimental measurements were
\(\reepoe = -0.010 \pm 0.021\) \cite{banner1972} and
\(\reepoe = 0.000 \pm 0.020\) \cite{holder1972},
consistent with zero.
The errors were dominated by the number of $\klpiopio$ events, 
less than 200 events in each experiment.
In a review\cite{kleinknecht}, Kleinknecht wrote that
\begin{quotation}
“It is not easy to improve substantially the experimental precision. A decision between the Superweak and milliweak models of \cp\ violation will therefore probably have to come from other experimental information outside the $K^0$ system.”
\end{quotation}
The problem was that the estimated value of 
$\reepoe$ based on the Kobayashi-Maskawa model was extremely small,
such as
\(<22 \times 10^{-4}\) \cite{ellis1976}.
To measure such a small effect, 
the $\reepoe$ had to be measured with a precision of $O(10^{-4})$,
which means that
more than $O(10^{6})$ $\klpiopio$ events (four orders of magnitudes larger than the statistics available at that time) had to be collected, 
and systematic uncertainties on $R$ had to be controlled to the level of $< 0.1\%$.
Increase in statistics had to wait for accelerators with higher energy and
 intensity, and detector technologies that could handle high event rates.

One large issue was systematic uncertainties.
For example, the branching fraction of $\klpippim$ can be measured by 
\begin{eqnarray}
	BR(\klpippim) & = & \frac{\Nobs{\klpippim}}{\NKLDecay\ \Acc{\klpippim}}\ ,
	\label{eq:br}
\end{eqnarray}
where 
$\Nobs{\klpippim}$ is the number of observed $\klpippim$ events,
$\NKLDecay$ is the number of $K_L$ decays, and
$\Acc{\klpippim}$ is the probability (acceptance) to observe the 
$\klpippim$ decay events which should be estimated by Monte Carlo simulations.
If the $\klpippim$ and $\klpiopio$ decays were collected in the same period, 
then $\NKLDecay$ would cancel in Eq.~(\ref{eq:R}).
Similarly, if $K_S$ decays were observed in the same period, 
Eq.~(\ref{eq:R}) becomes
\begin{eqnarray}
	R & = & \frac{\Nobs{\klpippim}\ \Nobs{\kspiopio}}{\Nobs{\kspippim}\ \Nobs{\klpiopio}}
		\cdot
			\frac{\Acc{\kspippim}\ \Acc{\klpiopio}}{\Acc{\klpippim}\ \Acc{\kspiopio}} ,
\end{eqnarray}
where $\Nobs{x}$ and $\Acc{x}$ are the number of observed events and the acceptance, respectively,
for the decay $x$.
However, $\Acc{\klpippim}$ and $\Acc{\klpiopio}$ are different because the 
final state particles are detected with different detectors.
The
$\Acc{\klpippim}$ and $\Acc{\kspippim}$ are also different because the 
decay position distributions are different due to the life time difference between 
$K_L$ and $K_S$, 
and the acceptance depends on the decay position.
The key of the experiments to measure $\reepoe$ was thus to 
understand 
the ratios of the four acceptances to the level of $<0.1\%$.

\subsection{Experiments in 1980's}
In 1980's, there were two experiments which measured $\reepoe$;
NA31 at CERN, and E731 at Fermilab.

\begin{figure}[htbp]
	\centering
	\includegraphics[width=0.8\linewidth]{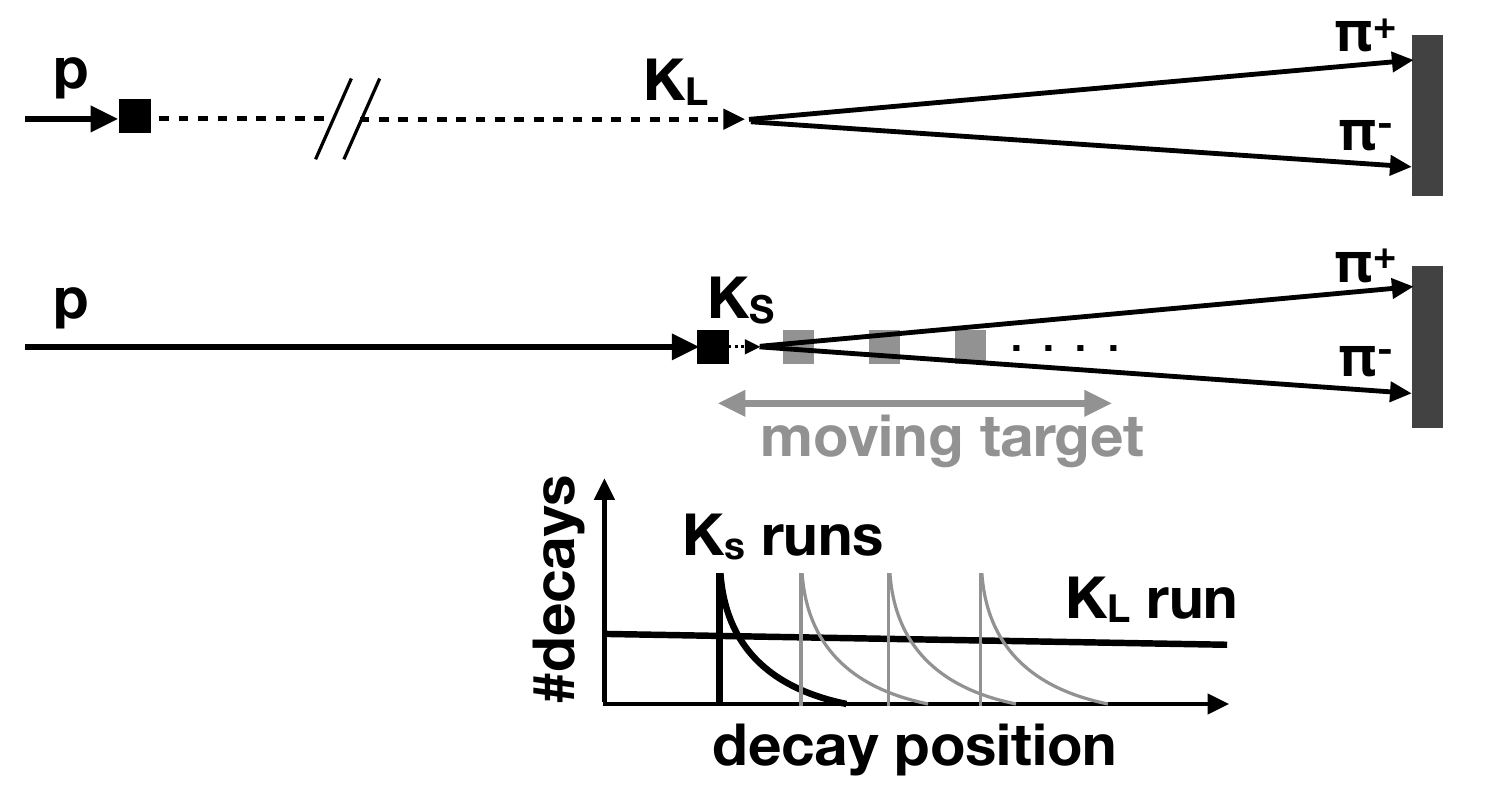}
	\caption{Schematic view of the $K_L$ (top) and $K_S$ (middle) runs 
		in CERN NA31.  
		By moving the $K_S$ production target, 
		decay position distributions for $K_S$ covered the $K_L$ decay 
		positions (bottom).
	}
	\label{fig:na31}
\end{figure}

The NA31 experiment collected $\pippim$ and $\piopio$ events simultaneously, 
and $K_L$ and $K_S$ runs separately.
The four photons from the $\piopio$ decays were detected by a liquid Argon calorimeter.
The energies and positions of the charged pions from the $\pippim$ decays 
were measured with a hadronic calorimeter.
To make the final state topology 
similar to that of $\piopio$ events,
a magnetic spectrometer was not used. 
To mimic the long decay position distribution of $K_L$, 
for the $K_S$ run, a target to produce $K_S$ was inserted and 
moved along the beam line inside the decay region,
as shown in Fig.~\ref{fig:na31}.
The events from the four decay modes were grouped by kaon energy and decay position bins.
The double ratio $R$ was calculated in each bin and then averaged.

\begin{figure}[htbp]
	\centering
	\includegraphics[width=0.8\linewidth]{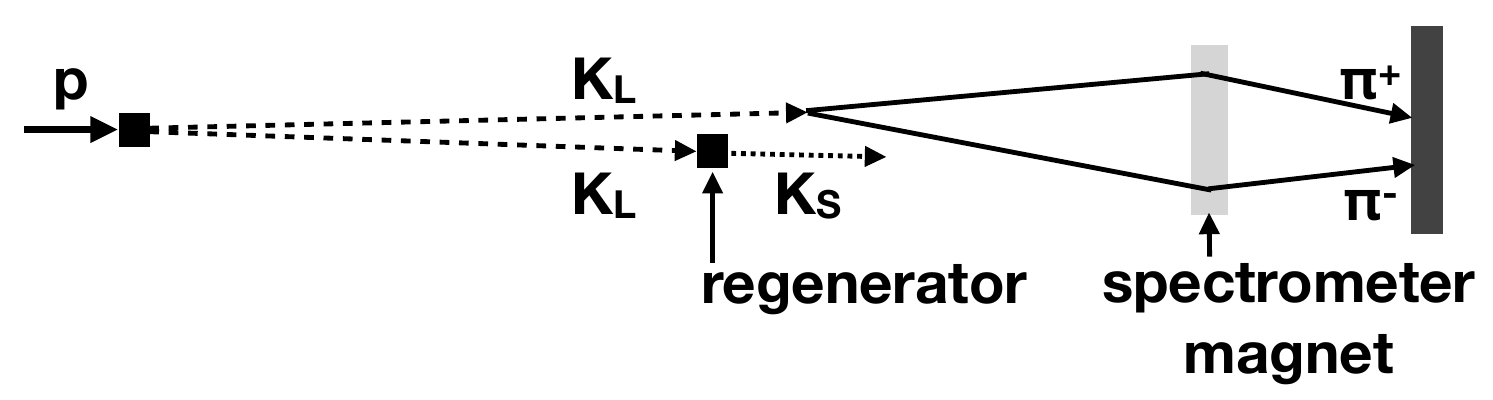}
	\caption{Schematic view of the $K_L$ and $K_S$ beams in Fermilab E731,
		created by placing a regenerator in one of the two $K_L$ beams.}
	\label{fig:e731}
\end{figure}

The E731 experiment collected the $K_L$ and $K_S$ decays simultaneously.
By placing a material in one of the two $K_L$ beams 
as shown in Fig.~\ref{fig:e731}
to regenerate $K_S$%
\footnote{%
Because materials are made of quarks, $\kobar$ containing $s$ quark has a
higher cross section than $\ko$ containing $\bar{s}$ quark, to conserve the baryon number.
Thus if 
\(\ket{K_L} \sim \ket{\kodd} \propto \ket{\ko} - \ket{\kobar}\)
passes through a material, and the wave function becomes
\(\ket{\psi} = (a\ket{\ko} - b\ket{\kobar})/\sqrt{2} = (a+b)/\sqrt{2}\ \ket{K_L} + (a-b)/\sqrt{2}\  \ket{K_S}\), the $K_S$ component appears.},
E731 effectively created $K_L$ and $K_S$ beams 
with a fixed $\Nobs{K_L}/\Nobs{K_S}$ ratio.
The $K_L$ and $K_S$ events were identified by reconstructing the event and
finding in which beam the kaon decayed.
Note that because charged pions and photons from the decays
spread out, the detectors themselves were 
insensitive to in which beam the parent $K$ decayed.
Initially, $\pippim$ and $\piopio$ events were collected separately, 
but later, all four decay modes were collected simultaneously.
To confirm that the acceptances were understood, the decay position distributions of 
high-statistics samples of \(K_L \to \pi^\pm e^\mp \nu\) and 
\(K_L \to \pi^0\pi^0\pi^0\) decays were compared between data and MC.

The NA31 measured
\(\reepoe = (23 \pm 6.5) \times 10^{-4}\), $3.5\sigma$ away from zero \cite{barr1993}, 
whereas the E731 measured
\(\reepoe = (7.4 \pm 6.0) \times 10^{-4}\), consistent with zero with $1.2\sigma$ \cite{gibbons}.
Many critical questions on various systematic effects were raised between
the groups, but none of them could resolve the difference in the results.
At the end, the both groups decided to 
build new experiments with an order of magnitude higher precision.

\subsection{Final experiments in 1990's}
Fermilab KTeV E832 experiment 
pursued the double $K_L$ and $K_S$ beam technique and 
collected the four decay modes simultaneously.
A new high intensity beam line with a better collimation scheme 
was built to increase the kaon rates and to reduce accidental hits in the detectors.
The electromagnetic calorimeter was newly built with 3100 CsI crystals 
with improved energy and position resolutions
to reduce systematic uncertainties.
The waveforms of the calorimeter signals were digitized at 53 MHz and recorded.
A new fast data acquisition system allowed to collect high statistics sample of 
\(K_L \to \pi e \nu\) and \(K_L \to 3\pi^0\) events to study systematic effects,
in addition to the $K \to \pi\pi$ events.

CERN NA48 experiment introduced a new technique to 
collect $K_L$ and $K_S$ events.
A small fraction of protons passing through the $K_L$ production target 
were guided to another production target far downstream to
produce $K_S$.
The $K_L$ and $K_S$ beams were designed to overlap in the detector region 
to reduce systematic uncertainties.
The events whose timings were correlated with the timings of the protons 
hitting the $K_S$ production target were identified as $K_S$.
The momentum of charged pions were measured with a magnetic spectrometer this time.
A new Liquid Krypton calorimeter was built to improve the photon energy resolution.

\begin{figure}[htbp]
	\centering
	\includegraphics[width=\linewidth]{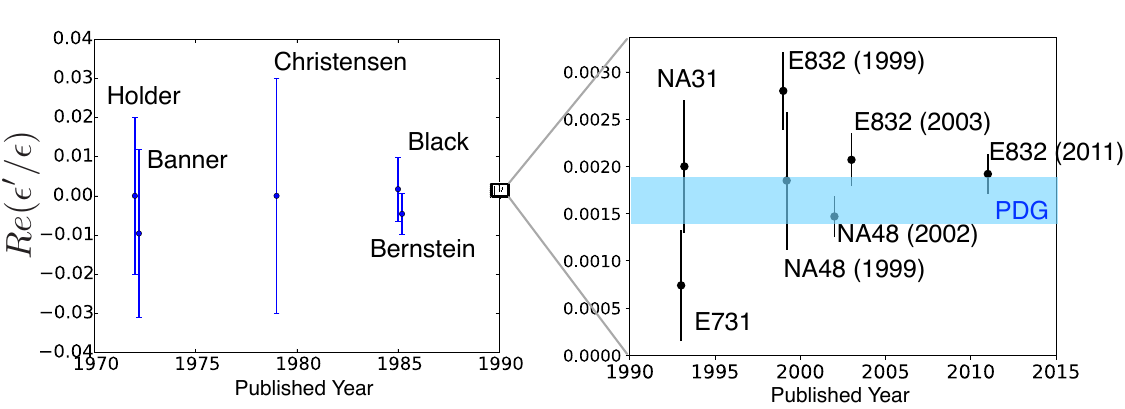}
	\caption{History of $\reepoe$ as a function of the published year.
		Labels on data points show the first author or the experiment name
		with optional published year in parenthesis.  
		The band on the right plot shows the average value by the Particle Data Group\cite{pdg_epoe}.}
	\label{fig:epoe_history}
\end{figure}

In 1999, with partial dataset, the Fermilab KTeV E832 experiment published 
\(\reepoe = (28.0 \pm 4.1) \times 10^{-4}\)\cite{ktev_1999}, 7$\sigma$ away from zero, and
CERN N48 also published 
\(\reepoe = (18.5 \pm 7.3) \times 10^{-4}\)\cite{na48_1999}, both
before \cp\ violation in B-meson system was first observed in 2001
\cite{babar_cp, belle_cp}.
At the end, the combined result 
based on full datasets of both experiments
was 
\(\reepoe = (16.8 \pm 1.4) \times 10^{-4}\) \cite{ktev_2011},
and the PDG's fit result is 
\(\reepoe = (16.6 \pm 2.3) \times 10^{-4}\) \cite{pdg_epoe}.
The number of $\klpiopio$ were 
\(6 \times 10^6\) for KTeV and
\(1.5 \times 10^6\) for NA48.
This put an end to the long awaited question.
The kaon experiments rejected \superweakModel\ as a sole source of \cp\ violation, and
supported the Kobayashi-Maskawa model.
Figure \ref{fig:epoe_history} shows how the measured $\reepoe$ has 
improved over the years.
The experimental results from $K$ and $B$ mesons 
gave solid foundations for the Kobayashi-Maskawa model 
to be a part of the \sm.

\section{Search for CP violation caused by new physics}
After the Kobayashi-Maskawa model was established,
kaon experiments moved on to search for new physics beyond the \sm\ 
that violates the \cp\ symmetry,
because the \cp-violation mechanism in the \sm\ is not large enough to 
explain the matter-antimatter asymmetry of the universe.

In principle, new physics could affect $\reepoe$, 
but it is difficult to observe it at this moment.
The amplitudes of the gluon penguin and $Z^0$ penguin diagrams
cancel each other, and lattice calculations for the gluon penguin diagram
need more years to give accurate estimate on $\reepoe$ even for the \sm\ alone.

\begin{figure}[htbp]
	\centering
	\includegraphics[width=\linewidth]{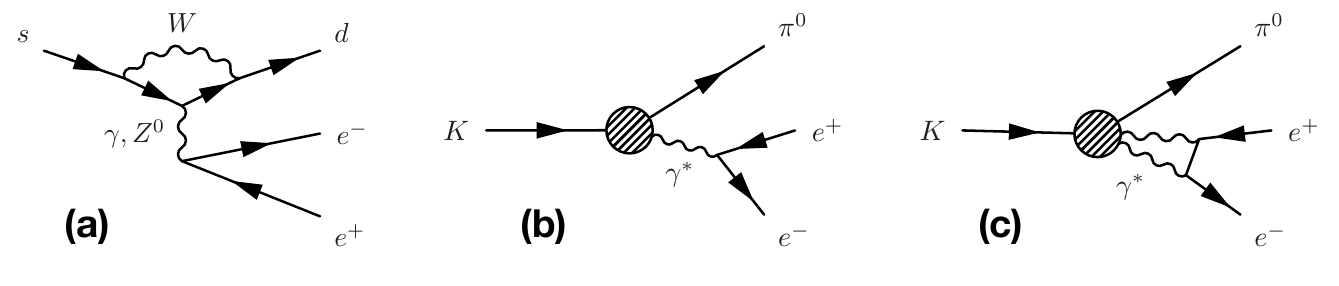}
	\caption{Feynman diagrams of the \(K_L \to \pi^0 e^+ e^-\)  decay;
	(a): penguin diagram and (b), (c): long distance 
		contributions with virtual photons.
		}
	\label{fig:pi0ee_diagrams}
\end{figure}

One way to search for such new physics 
is to use rare \cp-violating decays.  
If the contribution of the \sm\ is small, 
a contribution of new physics to the decay would be more visible.
The \(K_L \to \pi^0 e^+ e^-\) and \(K_L \to \pi^0 \mu^+ \mu^-\) decay modes 
are such candidates, because as shown in Fig.~\ref{fig:pi0ee_diagrams} (a), 
it has no gluon contribution.
However, contributions of decays with virtual photons 
shown in Fig.~\ref{fig:pi0ee_diagrams} (b) and (c) should be understood.

\begin{figure}[htbp]
	\centering
	\includegraphics[width=0.3\linewidth]{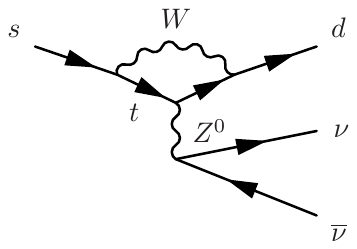}
	\caption{Penguin diagram for the $\klpionn$ decay in the \sm.}
	\label{fig:pi0nn_penguin}
\end{figure}

The $\klpionn$ decay is free from such virtual photon contributions, as shown in Fig.~\ref{fig:pi0nn_penguin}.
In the \sm, 
the decay amplitude of the $\kppipnn$ is governed by 
$|V_{td}|$, whereas 
the decay amplitude of $\klpionn$, $A(\klpionn)$, is proportional to 
\(Im(V_{td})\), as
\begin{eqnarray}
	A(\klpionn) & \propto & A(\ko \to \pionn) - A(\kobar \to \pionn)\\
			& \propto & V_{td} - V_{td}^*\\
			& \propto & Im(V_{td})
			\label{eq:Aklpionn} .
\end{eqnarray}
The expected branching fractions based on the \sm\ and
currently known CKM parameters are
\(BR(\klpionn) = (2.94 \pm 0.15) \times 10^{-11}\) and 
\(BR(\kppipnn) = (8.60 \pm 0.42) \times 10^{-11}\)\cite{buras2203}.
If the measured branching fractions of $\kpinn$ are different from the 
\sm\ predictions, it signifies the existence of new physics
as pointed out by Littenberg\cite{littenberg}.
Various new physics scenarios covering unexplored branching fraction region 
are reviewed in \cite{buras1507}.

Because the predicted branching fractions are small, 
the keys of the $\kpinn$ decay experiments are to produce a large number of kaons 
and to suppress background events.

\subsection{$\kppipnn$}
Major backgrounds  for the $\kppipnn$ decay are 
the $\kmutwo$ decay in which $\mu^+$ is misidentified as $\pi^+$, and 
the $\kppippio$ decay in which the two photons from the $\pi^0$ escaped detection.
 
One of the early experiments that searched for the $\kppipnn$ decay 
was KEK E10.
The charged kaons were stopped in a target,
and the decayed $\pi^+$ was identified by its path length in detectors, and 
observing the \(\pi^+ \to \mu^+ \to e^+\) decay chain.
The $\kppippio$ background was suppressed by detecting photons 
with lead glass blocks located on the other side of charged pion tracking detectors.
The experiment set an upper limit on the branching fraction as
\(BR < 1.4 \times 10^{-7}\) (90\% CL) \cite{asano}.

To increase the acceptance of the $\kppipnn$ decay, 
BNL E787/E949 experiments 
surrounded the $K^+$ stopping target with 
a cylindrical spectrometer and range counters to measure
the momentum and energy of $\pi^+$, respectively.
The waveforms in the range counters were recorded
to track the \(\pi^+ \to \mu^+ \to e^+\) decay chain to  
identify $\pi^+$.
The target was also surrounded by photon veto counters to
suppress $\kppippio$ background.
At the end, BNL E787/E949 measured
\(BR = (1.73^{+1.15}_{-1.05}) \times 10^{-10}\)\cite{e949}
based on
7 observed events where 0.93 background events were expected.

To increase the number of events by an order of magnitude, 
experiments proposed later chose to use high-momentum $K^+$'s decaying in flight.
By not having a stopping target in a high intensity beam, 
one can avoid producing secondary particles in the target.
Also, vetoing the $\kmutwo$ background is easier because 
high-momentum muons can easily penetrate through materials
whereas charged pions cannot:
this removed the need to track
the \(\pi^+ \to \mu^+ \to e^+\) decay chain 
for microseconds.
The $\kppippio$ background was suppressed by vetoing high energy photons 
decaying toward downstream.
Another background is caused by $\pi^+$ scattered from the beam.
Earlier stopping $K^+$ experiments suppressed $\pi^+$'s with a mass separator, 
but the same technique cannot be used for high momentum $K^+$'s and $\pi^+$'s.

Fermilab CKM experiment planned to apply high frequency electric fields at two locations
on a momentum-selected beam 
to kick out $\pi^+$'s which fly with a speed different from $K^+$'s.
The experiment was however, canceled during preparation.

CERN NA62 experiment decided not to remove $\pi^+$'s from the beam, 
but to identify $K^+$ with a fast differential Cherenkov counter.
Also, the momentum, direction, and timing of each $K^+$ is 
measured with pixel detectors and dipole magnets.
The decayed $\pi^+$ is identified by a ring-imaging Cherenkov counter,
and its momentum is measured with a magnetic spectrometer in vacuum.
The background from $\kmutwo$ decays is suppressed by identifying muons 
with NA48's Liquid Krypton calorimeter and scintillators between and behind iron plates.
The background from $\kppippio$ is suppressed by detecting photons from the decay with 
ring-shaped lead glass modules at multiple locations in the decay region, 
a liquid Krypton calorimeter placed downstream, and lead/scintillator calorimeter in the beam. 
In the data sample collected in 2016-2018, 
NA62 observed 20 events where 7 background events were expected, 
and measured
\(BR = (1.06^{+0.40}_{-0.34} \pm 0.09) \times 10^{-10}\) \cite{na62_jhep06}.
NA62 will improve the sensitivity with more data taken with improved beam line 
and detectors.

\subsection{$\klpionn$}
For the $\klpionn$ decay, the major issues is
how to identify the decay because only visible particles are two photons from the $\pi^0$.
The major background is the $\klpiopio$ decay where two of the four photons
in the final state escaped detection, 
but there are also many unexpected backgrounds because observables are limited.

The very first upper limit on the branching fraction was set as
\(BR < 7.6 \times 10^{-3}\) (90\% CL)\cite{littenberg}
by using the photon energy spectrum measured in an old experiment on 
$\klpiopio$.
Fermilab KTeV E799-II experiment had a one-day special run to
collect two-photon events, and set
\(BR < 1.6 \times 10^{-6}\) (90\% CL) \cite{ktev2g}.
Later, KTeV E799-II searched for the decay by using the
\(\pi^0 \to e^+ e^- \gamma\) decay to identify the $\pi^0$, 
reconstruct the decay vertex and the transverse momentum of $\pi^0$,
and set
\(BR <  5.9 \times 10^{-7}\) (90\% CL) \cite{kteveeg}.
Although the technique using the \(\pi^0 \to e^+ e^-\gamma\) decay is cleaner, 
it is less sensitive than the technique using the \(\pi^0 \to \gamma \gamma\) decay because of 
the small branching fraction of the \(\pi^0 \to e^+ e^- \gamma\) decay 
(1.2\%), and a small acceptance for detecting 
$e^+e^-$ pairs with a small opening angle.
Later experiments thus chose to use \(\pi^0 \to \gamma \gamma\) to search for the 
$\klpionn$ decay.

KEK E391a using 12 GeV protons 
was the first experiment dedicated for
the $\klpionn$ decay.
To suppress the background caused by the
$\klpiopio$ decay with two missing photons, 
and to collect events with only two photons, 
the decay region was surrounded by a hermetic photon veto detector system and
a CsI calorimeter covering downstream.
To avoid a background caused by a photon lost in a beam pipe, 
the beam pipe was removed and 
the entire photon veto system and the calorimeter were placed inside 
a vacuum tank, instead.
The decay vertex ($Z$) was reconstructed by assuming that the two photons 
were originated from a $\pi^0$.
The transverse momentum of $\pi^0$ ($P_T$) was then reconstructed based on
the decay vertex and the energies and hit positions of the two photons
assuming that the $K_L$ decayed at the center of the beam line.
A narrow $K_L$ beam was made to have a small $P_T$ resolution while
keeping the necessary $K_L$ yield which is proportional to the beam size.
Because some momentum was carried away by neutrino pairs, $P_T$ 
was required to have a finite value.
E391a set 
\(BR < 2.6 \times 10^{-8}\) (90\% CL) based on no observed events 
in the signal region defined in a \(P_T - Z\) plane\cite{e391a}.

As the construction of the J-PARC accelerator facility began, 
KOTO experiment was built to utilize its intense slow-extracted 30-GeV proton beam.
A new narrow $K_L$ beam line with a clean collimation scheme was built.
The E391a's vacuum tank and cylindrical photon veto counters were moved to J-PARC.
The electromagnetic calorimeter was replaced with 
2716 CsI crystals originally used in FNAL KTeV
to improve 
the detection efficiency and shower shape measurements.
A new data acquisition system 
records waveforms from all the detectors channels at 125 MHz (some at 500 MHz) to identify overlapping pulses.

Many unforeseen backgrounds were encountered as the sensitivity increased, 
and they were solved one by one.
One new background was caused by a neutron hitting the calorimeter
and producing another neutron, making two clusters.
This background was suppressed 
by installing thin photosensors on the upstream surface of the calorimeter to 
select photons which interact near the upstream end.
Another background was found to be 
caused by a small contamination ($O(10^{-5})$) of $K^\pm$ in 
the beam decaying into \(\pi^0 e^\pm \nu\) with undetected $e^\pm$
\cite{koto2016_2018}.
This was suppressed by installing a thin scintillation detector in the beam, 
and later, by installing another sweeping magnet in the beam.
The current best limit set by KOTO is 
\(BR < 3.0 \times 10^{-9}\) (90\% CL) \cite{koto2015}.
The experiment continues to collect more data with
improved detectors and higher beam power
to search for an order of magnitude enhancement on the branching fraction
by new physics.

A new experiment called KOTO II is being planned
to search for a smaller enhancement by new physics 
by increasing the sensitivity by more than two orders of magnitude.
To increase the $K_L$ yield, $K_L$ will be extracted at a smaller angle
($5^\circ$, compared 16$^\circ$ at KOTO) from the proton beam.
To increase the acceptance of the signal events, 
the decay region will be extended from 2 m to 12 m,
and the calorimeter will be enlarged from 2 m to 3 m in diameter.
In 36 months of running, 
the experiment expects to observe 35 \sm\ signal events on top of 
56 background events with the 4.7$\sigma$ significance\cite{koto2}.
If the measured branching fraction deviates from the \sm\ prediction by 44\%
due to new physics, 
it can be claimed with the 90\% confidence level.

\begin{figure}[htbp]
	\centering
	\includegraphics[width=0.9\linewidth]{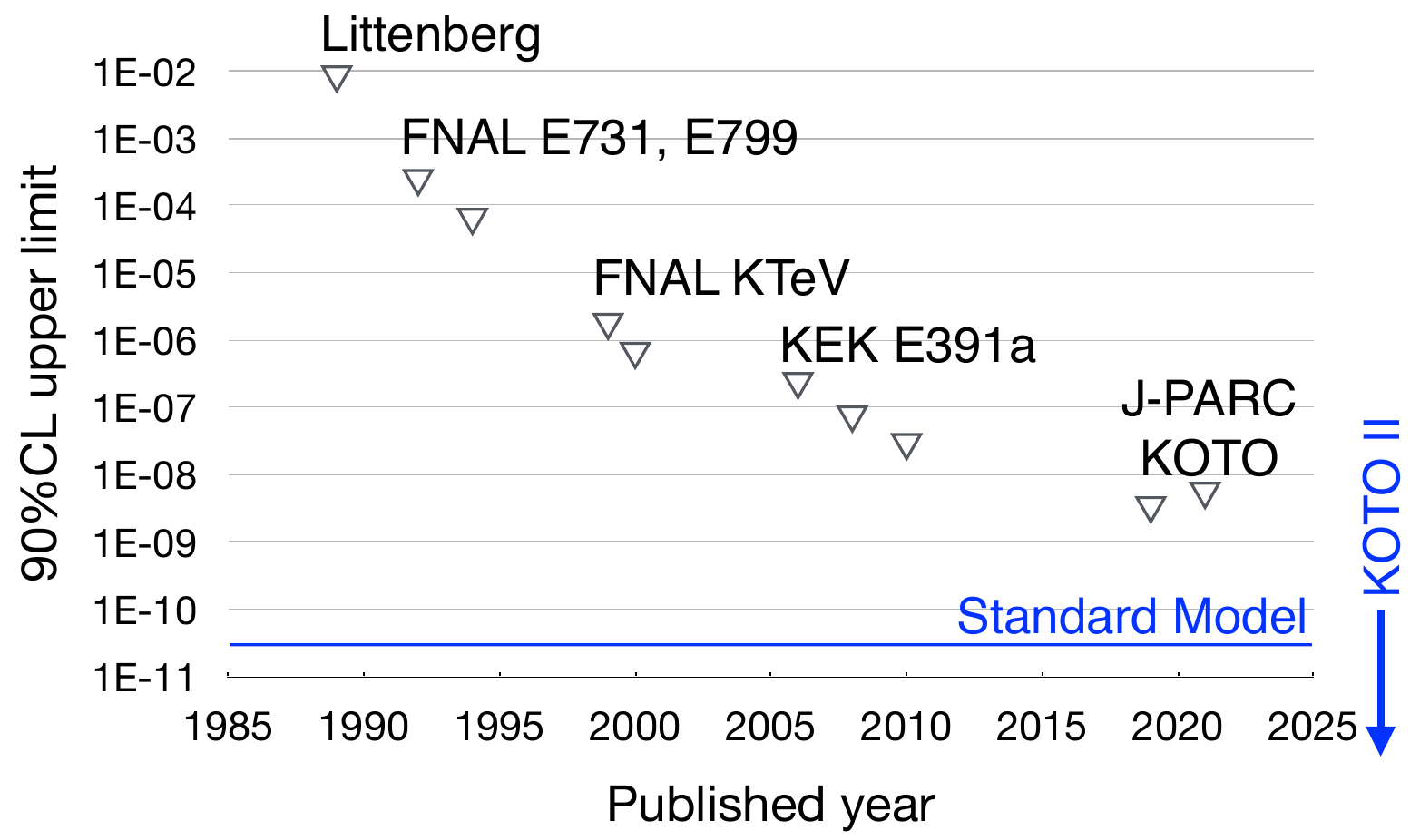}
	\caption{History of upper limits of the branching fraction of $\klpionn$
			shown as a function of the publication year.}
	\label{fig:pi0nn_year}
\end{figure}

Since 1989, the upper limit on the branching fraction of $\klpionn$ has been 
reduced by 5 orders of magnitudes, 
as shown in Fig.~\ref{fig:pi0nn_year}, and
KOTO II is being prepared to observe the events.

\section{Concluding remarks}
Neutral kaon is a unique system to study \cp\ violation,
because it becomes a nearly \cp-odd state just by 
placing detectors away from a production target.
Kaon experiments studying \cp\ violation have improved 
over time,
thanks to 
new accelerators with higher intensities, 
new detector technologies with higher sensitivities and rate capabilities,
and many ingenious ideas.
In 2073, at a symposium celebrating the 100th anniversary, 
I hope somebody will give a review talk starting as
\begin{quotation}
``Back in 2020's, the Kobayashi-Maskawa model was the only source of 
\cp\ violation, and they were still trying to find $\klpionn$.
Now we see thousands of those events.
What made this possible?''
\end{quotation}

\section*{Acknowledgment}
I would like to thank the organizers for inviting me to give a talk at the
honorable symposium,
\textit{Accomplishments and Mysteries in Quark Flavor Physics $\sim$ 50th Anniversary of Kobayashi-Maskawa Theory (KM 50)}
which was held on Feb.~9-10, 2023 at KEK.

%

%

\vspace{0.2cm}


\let\doi\relax


\newcommand{\paper}[5]{#1, #2 \textbf{#3}, #4 (#5).}
\newcommand{\prd}{Phys.\ Rev.\ D}
\newcommand{\prl}{Phys.\ Rev.\ Lett.}
\newcommand{\pl}{Phys.\ Lett.}
\newcommand{\plb}{Phys.\ Lett.\ B}
\newcommand{\ptp}{Prog.\ Theor. Phys.}
\newcommand{\ptep}{Prog.\ Theor.\ Exp.\ Phys.}
\newcommand{\epjc}{Euro.\ Phys.\ J.\ C}
\newcommand{\jhep}{J.\ High Energ Phys.}

\end{document}